\documentclass[twocolumn,aps,prl,superscriptaddress,showpacs,preprintnumbers,amsmath,amssymb]{revtex4}

\usepackage[dvips]{graphicx}
\usepackage{dcolumn}

\begin{document}
\title{Upper Limit on Gravitational Wave Backgrounds at 0.2 Hz with Torsion-bar Antenna}

\author{Koji Ishidoshiro}
\altaffiliation[Present address: ]
{High Energy Accelerator Research Organization, Tsukuba, Ibaraki 305-0801, Japan}
\email{koji@post.kek.jp}
\affiliation{Department of Physics, University of Tokyo, 
7-3-1 Hongo, Bunkyo-ku, Tokyo 113-0033, Japan}

\author{Masaki Ando}
\affiliation{Department of Physics, Kyoto University, Kyoto 606-8502, Japan}

\author{Akiteru Takamori}
\affiliation{Earthquake Research Institute, University of Tokyo,
Bunkyo-Ku, Tokyo 113-0032, Japan}

\author{Hirotaka Takahashi}
\affiliation{Earthquake Research Institute, University of Tokyo,
Bunkyo-Ku, Tokyo 113-0032, Japan}
\affiliation{Department of Humanities, Yamanashi Eiwa College, 
Kofu, Yamanashi 400-8555, Japan}

\author{Kenshi Okada}
\affiliation{Department of Physics, University of Tokyo, 
7-3-1 Hongo, Bunkyo-ku, Tokyo 113-0033, Japan}

\author{Nobuyuki Matsumoto}
\affiliation{Department of Physics, University of Tokyo, 
7-3-1 Hongo, Bunkyo-ku, Tokyo 113-0033, Japan}

\author{Wataru Kokuyama}
\affiliation{Department of Physics, University of Tokyo, 
7-3-1 Hongo, Bunkyo-ku, Tokyo 113-0033, Japan}

\author{Nobuyuki Kanda}
\affiliation{Department of Physics, Osaka City University, Osaka 558-8585, Japan}

\author{Yoichi Aso}
\affiliation{Department of Physics, University of Tokyo, 
7-3-1 Hongo, Bunkyo-ku, Tokyo 113-0033, Japan}

\author{Kimio Tsubono}
\affiliation{Department of Physics, University of Tokyo, 
7-3-1 Hongo, Bunkyo-ku, Tokyo 113-0033, Japan}

\date{\today}

\begin{abstract} 
We present the first upper limit on gravitational wave (GW) backgrounds at an unexplored frequency of 0.2~Hz 
using a torsion-bar antenna (TOBA). 
A TOBA was proposed to search for low-frequency GWs. 
We have developed a small-scaled TOBA 
and successfully found $\Omega_{\rm gw}(f) < 4.3 \times 10^{17}$ 
at 0.2~Hz as demonstration of the TOBA's capabilities, 
where $\Omega_{\rm gw}(f)$ is the GW energy density per logarithmic 
frequency interval in units of the closure density.  
Our result is the first nonintegrated limit to bridge the gap 
between the LIGO band (around 100~Hz) and the Cassini band ($10^{-6}-10^{-4}$~Hz). 
\end{abstract}


\maketitle
\textit{Introduction.}--\\
The standard cosmology predicts not only the cosmic microwave background (CMB) from the last scattering surface, 
but also gravitational wave backgrounds (GWBs) as the result of processes that take place very shortly after the big bang \cite{maggiore}. 
Astrophysical GWBs could also emerge from the superposition of a large number of unresolved sources \cite{regimbau}. 
The proper frequency characteristic of the cosmological GWBs is determined by their generation mechanisms  
and the state of the Universe when the wavelength of the GWBs crossed the Hubble horizon. 
The frequency characteristic of the astrophysical GWBs depends on the motion scale of GW sources. 
Measurements of GWBs in various frequency bands enable us to separate the GWBs according to their origins, 
and then reveal how the Universe evolved from its very early epoch.  
Therefore, GWB detection and characterization is one of the greatest challenges
in not only GW experiments, but also
cosmology and astronomy. 
 
A number of experiments have been performed to detect GWBs 
or constrain $\Omega_{\rm gw}(f)$ at several frequencies, 
where $\Omega_{\rm gw}(f)$ is the cumulative energy density of GWBs per unit logarithmic frequency, 
divided by the critical energy density to close the Universe. 
At around 100 Hz, the Laser Interferometer
Gravitational Wave Observatory (LIGO) has found $\Omega_{\rm gw}(f)<6.9\times
10^{-6}$ at the 95\% confidence level \cite{ligo}. 
A pair of synchronous recycling interferometers has set $\Omega_{\rm gw}(f)<1.2\times
10^{26}$ for 100~MHz GWBs \cite{akutsu}. 
At 907~Hz, a cross-correlation measurement between the Explorer and Nautilus
cryogenic bar detectors has placed $\Omega_{\rm gw}(f)<120$ \cite{bar}. 
At $10^{-6} - 10^{-3}$~Hz, the Cassini spacecraft has
established an upper limit by using spacecraft Doppler tracking \cite{cassini}. 
Based on the fluctuations in the pulse arrival times from PSR B1855+09, 
an upper limit has been found in the frequency band $10^{-9}-10^{-7}$~Hz \cite{maggiore}. 
Measurement of the CMB at large angular scales 
indicates an upper limit at very low frequencies ($10^{-18} - 10^{-16}$~Hz) \cite{cobe}. 
In addition, the integrated $\Omega_{\rm gw}=\int \Omega_{\rm gw}(f) d(\ln f)$ is indirectly constrained by 
the helium-4 abundance resulting from big-bang nucleosynthesis (BBN) \cite{maggiore}  
and measurements of the CMB and matter power spectra \cite{cmb}.

Currently, the low-frequency band at the range of $0.01 - 1$~Hz is still unexplored, 
although several low-frequency GW antennas are being proposed and
developed, -e.g., the Laser Interferometer Space Antenna \cite{lisa}, 
DECi-hertz Interferometer Gravitational wave Observatory \cite{decigo}, and Atomic Gravitational wave Interferometric Sensor \cite{agis}. 

In the previous Letter, we have proposed a torsion-bar antenna (TOBA) 
to search for GWs at 0.01 - 1~Hz \cite{ando}. 
For example, it is realistic to achieve a GW strain-equivalent noise level  
of $\sim 10^{-18}$~Hz$^{-1/2}$ at 0.1~Hz, 
even with a ground-based configuration. 
In this configuration, 
the observable range reaches about 10~Gpc for $10^5$~M$_{\rm SUN}$ black hole events. 
To evaluate TOBA's capabilities for low-frequency GW observations, 
we developed a small-scaled TOBA and performed a short observational run. 
This Letter presents the first upper limit on GWBs at 0.2~Hz obtained from the small-scaled TOBA 
as demonstration of the capabilities, and discusses the future strategy of a TOBA.

\begin{figure}
 \includegraphics[width=8cm]{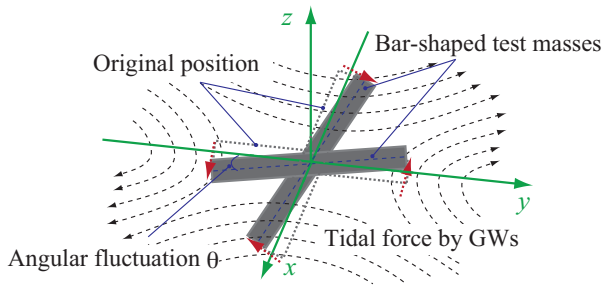}
 \caption{Principle of a TOBA. Two orthogonal test masses 
feel the tidal forces by incident GWs.}\label{fig1}
\end{figure}

\textit{TOBA.}--\\
A TOBA differs from conventional ground-based GW antennas in its fundamental sensitivity to GWs below 1 Hz. 
A TOBA consists of two rotational sensors and two bar-shaped test masses arranged parallel to 
the $x$-$y$ plane and orthogonal to each other (see Fig.~\ref{fig1}). 
Each test mass is suspended from its center; 
therefore, it behaves as a free mass in the rotational degree of freedom around the $z$ axis. 
When $\times$-polarized GWs of $h_{ij}(t)$ 
($h_{12}(t)=h_{21}(t)=h_{\times}(t)$ and $h_{ij}=0$ $((i,j)\neq (1,2), (2,1)))$ pass through a TOBA, 
tidal forces by the GWs will appear as angular fluctuations of the two test masses.

The angular fluctuation $\theta$ of a test mass from its original
position obeys the equation of motion \cite{ando}: 
\begin{equation}
 \tilde{\theta}(f)=\frac{q_{12}}{2I}\tilde{h}_\times(f) \hspace{0.4cm} \Big( f_{\rm res} \ll f \ll \frac{I}{2\pi}\gamma \Big), 
\end{equation}
where $I$, $q_{12}$ and $f_{\rm res}=1/(2\pi)\sqrt{I/\kappa}$ are 
the moment of inertia of the test mass, the dynamical quadrupole moment ($q_{12}=q_{21}$ and $q_{11}=-q_{22}$, see \cite{hirakawa}), and 
the rotational resonance frequency, respectively. 
Here, $\gamma$ and $\kappa$ are the damping constant and the spring constant around the $z$ axis, respectively. 
A tilde ($\verb|~|$) denotes the Fourier transformation. 

The equation of the motion of another test mass is also written as $\tilde{\theta}'=-q_{12}/(2I)\tilde{h}_\times(f)$, 
where $\theta'$ is the angular fluctuation of the test mass. 
The differential fluctuation $\Delta \theta(t)$ $(= \theta(t)-\theta'(t))$ is expressed as, 
\begin{equation}
\Delta \theta(f)=\frac{q_{12}}{I}\tilde{h}_\times(f) \hspace{0.4cm} \Big( f_{\rm res} \ll f \ll \frac{I}{2\pi}\gamma \Big). 
\end{equation}
GWs can be detected from $\theta$ with the single-mass configuration or 
$\Delta \theta$ with the differential-measurement configuration at $f_{\rm res} \ll f \ll I/(2\pi)\gamma$. 
In general, the resonance frequency $f_{\rm res}$ in a torsion pendulum is as low as a few mHz \cite{tp}. 
Thus, a TOBA can have a fundamental sensitivity in low frequency band (0.01 - 1 Hz).

\begin{figure}
 \includegraphics[width=8cm]{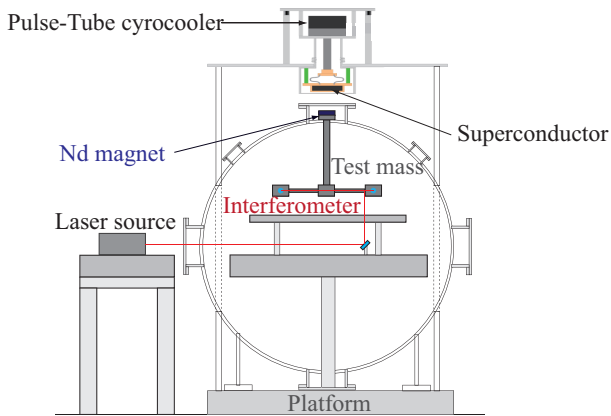}
 \caption{Conceptual design of small-scaled TOBA.}\label{fig2}
\end{figure}

\textit{Small-Scaled TOBA.}--\\
We developed a small-scaled TOBA as shown in Fig.~\ref{fig2}. 
The small-scaled TOBA mainly consists of a single test mass  
and a interferometric rotational sensor. As the first step, we applied the single-mass configuration, 
though we had to abandon the common mode noise rejection obtained from the differential measurement. 
Other unique equipment is a magnetic suspension based on the pinning effect of a 
type-II superconductor. This suspension can potentially provide large suspension forces with 
the low spring constant $\kappa$ and the low damping constant $\gamma$ 
without the drawbacks of contact \cite{maglev}. 

Our test mass has an inverted T-shape with a horizontal length of 22.5~cm, a vertical length of 19~cm, and a mass of 131~g. 
On the top of the test mass, a symmetric neodymium (Nd) magnet ($\phi$22~mm, t10~mm) is attached. 
The superconductor, which is made of a Gd-Ba-Cu-O compound and has 
a critical temperature of 92~K, is placed above the Nd magnet. 
When the superconductor is cooled to about 65~K by a
low-vibration pulse-tube cryocooler, the test mass is suspended by the magnetic forces 
between the Nd magnet and the superconductor. 
The shape of the test mass was chosen to bring the Nd magnet close to the superconductor and 
elongate the arms as much as possible. 
%
The spring constant and the damping constant of our suspension system are 
$\kappa= 3.6 \pm 2.1 \times 10^{-7}$~Nm/rad and 
$\gamma=1.2 \pm 0.7 \times 10^{-8}$~Nms/rad 
(a rotational resonance frequency of about 5~mHz and a quality factor of
about 1000), respectively. 
These values are similar to those of a typical tungsten-based torsion pendulum. 
The current damping constant is limited by the random collision of residual gas 
under $10^{-3}$~Pa condition.

We implement a laser Michelson interferometer using two mirrors attached on each end of the test mass. 
The angular fluctuation $\theta$ then is measured from the optical pass difference $\Delta l$ 
in the interferometer. 
The fluctuation $\theta$ is fed back to the test mass using coil-magnet actuators 
for the operation of the interferometer in its linear range. 
In this configuration, GW signals are read out from the feedback signal, just like the conventional antennas \cite{saulson}. 

As a laser source, a Nd:YAG laser is used with a wavelength of 1064~nm and an output power of 40~mW.
To compose the actuators with external coils, 
the test mass houses two samarium-cobalt (SmCo) magnets ($\phi$1~mm, t0.5~mm) at each end of the horizontal arm. 
The test mass and most of the readout interferometer are located
in a vacuum chamber, magnetic and thermal shields to avoid the effects of external disturbances. 

\begin{figure}
 \includegraphics[width=8cm]{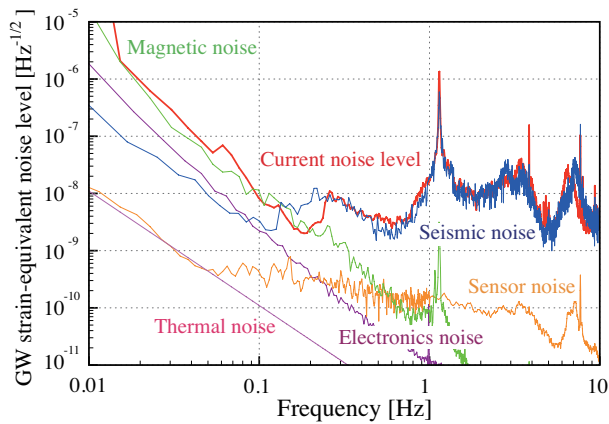}
 \caption{Measured GW strain-equivalent noise level. Estimated environmental noises 
(seismic noise and magnetic noise), technical noises (sensor noise and electronics noise) and thermal noise are also described.}\label{fig3}
\end{figure}

We find a GW strain-equivalent noise level $h(f)$ of $2\times 10^{-9}$~Hz$^{-1/2}$ 
around 0.2~Hz (Fig.~\ref{fig3}). 
Current noise level is limited by environmental disturbances: seismic noise above 0.2~Hz and magnetic noise below 0.2~Hz. 
The seismic noise is from the unexpected coupling 
between the translation of the test mass induced by seismic motion   
and $\Delta l$ in the interferometer. 
The magnetic noise is introduced by the coupling between the residual non-axial
symmetry of the Nd magnet and the fluctuation of external magnetic fields. 
Little differences of 
the structure between the measured noise level and the estimated environmental noises shown in Fig.~\ref{fig3} 
are due to measurement conditions. 
The peaks at 0.3~Hz, 1.2~Hz, 3.5~Hz and 7~Hz are identified 
as the effect of microseismic disturbances, the resonance of the rigid-body-pendulum of the test mass, 
the resonance of the platform of the chamber and the resonance of the lab's floor, respectively. 
Compression of the cryocooler also induces peaks at 3.9~Hz and 7.8~Hz. 

More details of instrumental status, including noise analysis and performance of the magnetic suspension, 
are published as individual papers \cite{ishidoshiro,takamori,ishidoshiro2}. 

\textit{Upper limit on GWBs.}--\\
On August 15, 2009, we performed a one-night observational run using the small-scaled TOBA. 
The total length of the data was about 7.5~hr. 

Using approximately $T_{\rm eff}=3.5$~hr of stable low-noise data, 
we place an upper limit on $\Omega_{\rm gw}(f_0)$ at $f_0=0.2$ Hz with a bandwidth of 10~mHz. 
Such arbitrary data selection is acceptable, since we assume stationary GWBs in this Letter. 

The data analysis pipeline consists of  
estimating a GW energy-equivalent spectral density $\Omega_{\rm eq}(f_0)$ and 
determining the upper limit $\Omega_{\rm gw}^{\rm UL}(f_0)$. 
Supposing that GWBs are isotropic and unpolarized, 
the $\Omega_{\rm eq}(f_0)$ can be related to $h(f_0)$: 
\begin{equation}\label{eqomega}
 \Omega_{\rm eq}(f_0)=\frac{10\pi^2}{3H_0^2}f_0^3h^2(f_0).
\end{equation} 
Here, $H_0$ is the Hubble constant, and 
its value is $70.4^{+1.3}_{-1.4}$ km/s/Mpc \cite{wmap7}. 

\begin{figure}
 \includegraphics[width=8cm]{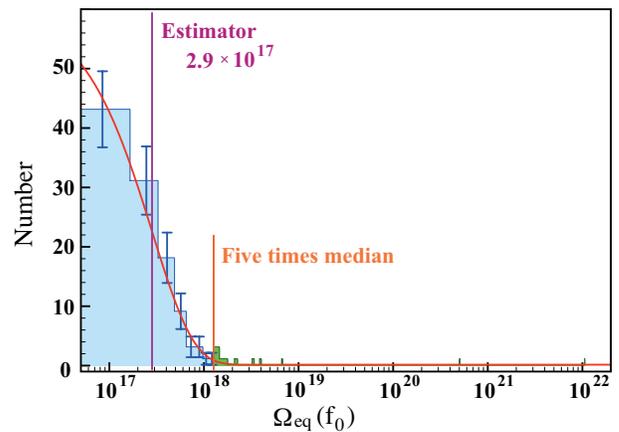}
 \caption{Distribution of $\Omega_{\rm eq}(f_0)$. 
Green boxes are rejected segments. 
Red curve is the result of the exponential fitting. 
Blue bars are Poisson errors of the distribution. 
}\label{fig4}
\end{figure}

To obtain the estimator of $\Omega_{\rm eq}(f_0)$, the data are divided into $t_{\rm s}=102.4$ sec segments. 
The length of the segment is selected such that 
the bandwidth is much smaller than the target frequency ($1/t_{\rm s} \ll f_0$)  
to obtain better frequency resolution and reduce the statistical error. 
We then have 120 ($=T_{\rm eff}/t_{\rm s}$) $\Omega_{\rm eq}(f_0)$ using Eq.~($\ref{eqomega}$). 
Then, the estimator $\bar{\Omega}_{\rm eq}(f_0)$ is $2.9^{+0.22}_{-0.24}\times 10^{17}$  
from the exponential fitting of the $\Omega_{\rm eq}(f_0)$ distribution, rejecting the segments 
whose $\Omega_{\rm eq}(f_0)$ are five times larger than 
the median calculated from all segments (see Fig.\ref{fig4}). The addition term is the statistical (fitting) error.

The upper limit on GWBs, $\Omega_{\rm gw}^{\rm UL}(f_0)$, 
can be described using a confidence level $C$ defined by, 
\begin{equation}
 C=\int^\infty_{\bar{\Omega}_{\rm eq}(f_0)}P(\Omega_{\rm es}(f_0)|\Omega^{\rm UL}_{\rm gw}(f_0))d\Omega_{\rm es}(f_0), 
\end{equation}
where $P(\Omega_{\rm es}(f_0)|\Omega^{\rm UL}_{\rm gw}(f_0))$ is a conditional probability distribution and 
obeys the Gaussian distribution:
\begin{equation}\label{eq:dist}
 P(\Omega_{\rm es}(f_0)|\Omega_{\rm gw}(f_0))\propto 
  \exp\Biggl[ -\frac{(\Omega_{\rm es}(f_0)-\Omega_{\rm gw}(f_0))^2}{2\Omega_{\rm gw}(f_0)^2/N} \Biggr].  
\end{equation}
Here, $\Omega_{\rm es} (f_0)$ is the estimator of $\Omega_{\rm gw}(f_0)$ using the $N(=109)$ samples. 
As a result we found $\Omega_{\rm gw}^{\rm UL}(f_0)=3.4^{+0.26}_{-0.29} \times 10^{17}$ at $C=0.95$. 

The biggest systematic error is the sensitivity from $\Delta l$ to $\theta$. 
It arises from the uncertainty of the beam spot positions in the end mirrors. 
From the size of the mirrors, we find that the maximum error is about $10\%$. 
For $\Omega_{\rm gw}^{\rm UL}(f_0)$, this error is about $20\%$, since $\Omega_{\rm eq}(f_0)$ 
is proportional to $\tilde{\theta}^2(f_0)$. 
Considering the statistical error and the systematic error,  
we finally obtain a conservative upper limit of $\Omega^{\rm UL}_{\rm gw} =4.3 \times 10^{17}$ at $C=0.95$.

\begin{figure}
 \includegraphics[width=8cm]{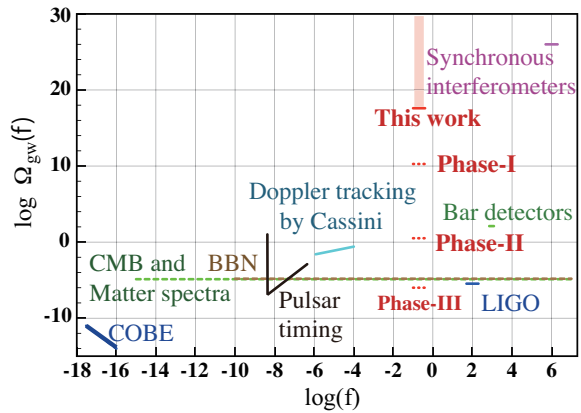}
 \caption{Upper limits $\Omega^{\rm UL}_{\rm gw}(f)$.  
 Red line is our new upper limit. 
 Red dotted lines are our expected limits in next three phases. 
 Current upper limits are also 
 described as solid lines \cite{maggiore, ligo, akutsu, bar, cassini,cobe}. 
 Dashed lines are the frequency-integrated upper limits \cite{maggiore, cmb}.
}\label{fig5}
\end{figure}

\textit{Discussion.}--\\
Comparison with the other experiments is shown in Fig.~\ref{fig5}. 
Using the small-scaled TOBA, we successfully found the first nonintegrated upper limit on GWBs 
at the unexplored frequency that bridges the gap between the LIGO band (around 100 Hz) and 
the Cassini band ($10^{-6}-10^{-4}$ Hz). Considering integrated upper limits,   
the energy density $\Omega_{\rm gw}(f)$ at 0.2~Hz is already constrained by 
the BBN \cite{maggiore} or the CMB measurements \cite{cmb}. 
However, astrophysical GWBs would be generated at much later times and, thus, would not be subject to the 
above limits. On the other hand, our limit can constrain such GWBs at 0.2 Hz. 
Therefore, our result complements other nonintegrated upper limits at 
different frequencies and integrated upper limits at 0.2~Hz. 

To built the TOBA with the final configuration \cite{ando}, we will have three phases.  
Applying well designed magnetic shields 
to the small-scaled TOBA, 
the initial phase (phase-I) aims at the thermal noise limited noise level ($h\sim 10^{-12}$~Hz$^{-1/2}$ at 0.1~Hz) 
under $10^{-7}$~Pa. Remember that the current thermal noise is limited by the residual gas noise.  
In this phase-I, we will move the small-scaled TOBA environmentally quieter site (Kamioka mine) 
and install the second test mass to obtain the common mode noise rejection for the seismic noise.   
The next phase (phase-II) is a middle-scaled TOBA using two 2~m scaled test masses. 
From the increase of the moment of inertia and the optimization of the mass-shape, 
force noises will be suppressed by a factor about 500. 
Target noise level is about $10^{-15}$~Hz$^{-1/2}$ at 0.1~Hz using locked Fabry-Perot interferometers as the rotational sensors. 
After that, we will be in final phase (phase-III).  
A major change from the previous phase is 10~m scaled and cryogenically cooled test masses. 
The cooling aims to suppress the thermal noises of the test masses and the suspensions. 
The TOBA with the final configuration will have $h\sim 10^{-18}$~Hz$^{-1/2}$ at 0.1~Hz. 
Estimated upper limits at each phases are described in 
Fig.~\ref{fig5} with a one-year observation by a pair of two TOBAs.  

\textit{Conclusion.}-- \\
A TOBA has been proposed 
to search low-frequency GWs  
even with a ground-based configuration. 
We have developed a small-scaled TOBA and placed the first nonintegrated upper limit on GWBs at 0.2~Hz. 
The new constraint is $\Omega_{\rm gw}(f)<4.3\times 10^{17}$ at 0.2~Hz with a bandwidth of 10~mHz. 
We experimentally demonstrated TOBA's capabilities and successfully opened the unexplored frequency band that the current GW antennas can not access.

This work was supported by a Grant-in-Aid for JSPS Fellows, Young Scientists~(A), 
GCOE for Phys.~Sci.~Frontier, MEXT, Japan. 
The authors thank I.~Buder for carefully reading the manuscript.

{}

\end{document}